# Impacts of National Cultures on Managerial Decisions of Engaging in Core Earnings Management

Muhammad Rofiqul Islam[1*] Abdullah Al Mehdi[2]

1. A.R. Sanchez, Jr. School of Business, Texas A&M International University, 5201 University Blvd, Laredo, TX 78041, USA

2. Starco Impex, Inc., 2710 S 11th St., Beaumont, TX 77701, USA

\* E-mail of the corresponding author: muhammadrofiqulislam@dusty.tamiu.edu

**Abstract**

This study investigates the impact of Hofstede's cultural dimensions on abnormal core earnings management in multiple national cultural contexts. We employ an Ordinary Least Squares (OLS) regression model with abnormal core earnings as the dependent variable. The independent variables analyzed include Hofstede's dimensions: Power Distance Index (PDI), Individualism (IDV), Masculinity (MAS), and Uncertainty Avoidance Index (UAI). Our findings reveal that individualism is positively associated with abnormal core earnings, suggesting that cultures characterized by high individualism may encourage practices that inflate earnings due to the prominence of personal achievement and rewards. In contrast, masculinity negatively correlates with abnormal core earnings, indicating that the risk-taking attributes associated with masculine cultures may deter earnings management. Interestingly, uncertainty avoidance is positively linked to abnormal core earnings, supporting the notion that managers tend to engage more in earnings management to minimize fluctuations in financial reports in cultures with high uncertainty avoidance. The relationship between power distance and abnormal core earnings is found to be non-significant, indicating no substantial effect in this context. These findings contribute to the literature on cultural influences in financial reporting, providing valuable insights for policymakers and multinational firms concerning the cultural contexts within which financial decisions and reporting occur.

## 1. Introduction

Despite extensive legislative measures aimed at reducing earnings management, company managers continue to participate in various types of earnings management. This has garnered the attention and concern of regulators and academics studying the practice (Healy & Wahlen, 1999; Leuz et al., 2003; Lopez & Rees, 2002). The increased regulatory focus on earnings management is a result of its effect on the accuracy and significance of accounting earnings figures (Ball & Brown, 2013; Beaver et al., 2012) and their impact on the pricing of debt and equity (Francis et al., 2004; Kothari, 2001). Given the connection between accounting figures and the valuation of securities, managers are motivated to manipulate accounting figures in a desired manner to deceive investors and other relevant stakeholders (such as regulators, suppliers, and customers). Unlike the exactitude of pure science, accounting permits the exercise of discretion and judgment when applying accounting principles and norms. Managers can select from many accounting systems and procedures, including depreciation methodologies, depreciation rates, and contingencies for uncollectible amounts. Managers can positively alter earnings figures by selecting a specific accounting system or approach instead of another. The act of altering financial figures to achieve desired earnings is called earnings management.

Research on the correlation between culture/value and earnings management is abundant, with a primary focus on two basic types of earnings management: real-earnings management and accrual-based earnings management. Several studies have examined the influence of country cultural differences on core earnings management, specifically focusing on the third category of earnings management. Research on the correlation between national cultures and core profits management was conducted within the framework of developed countries, hence constraining the applicability of such findings. This study seeks to fill the existing vacuum in the literature by investigating the influence of national cultural variations on core earnings management. The research will utilize a sample from a wide range of nations.

## 2. Literature Review

Extensive academic research has been conducted on earnings management, examining various aspects such as factors that predict, influence, and result from earnings management. This essay evaluates previous research on the correlation between national culture and earnings manipulation. The relationship between culture and earnings management can be understood through the lens of institutional theory, which posits that decision-makers' behavior is shaped by the shared understanding and value systems they absorb from their surroundings (Zucker, 1983). As an informal institution, culture influences individuals' motivation and reasoning for participating in specific activities or exhibiting certain organizational behaviors (North, 1990). Managers' judgments about earnings management are influenced by their value systems, beliefs, perceptions, and societal customs (Deephouse et al., 2016).

The theoretical justification for the connection between culture and profits management aligns with the Gray (1988) model, which elucidates the correlation between cultural dimensions and accounting values. Considering the theoretical context, the existing literature examined the correlation between national culture and profits management and identified multiple characteristics. Prior research has examined the correlations between several aspects of national culture, such as collectivism, uncertainty avoidance, masculinity, power distance (Hofstede & Hofstede, 2001), and the practice of earnings management. Furthermore, earnings management's impact on investor protection and the adoption of the International Financial Reporting System (IFRS) are influenced by national culture. Furthermore, national culture influences how earnings management is perceived and accepted.

### *2.1 National Culture and Earnings Management*

According to Gray (1988), cultural features play a crucial role in shaping accounting systems, professional standards, disclosure requirements, and evaluating financial management procedures inside a company. The author provided empirical evidence to support the assertion that national cultural characteristics significantly influence accounting values. An organization's Earnings management decisions entail assessing many options, and decision-makers select alternatives based on their comprehension, convictions, and principles. National culture encompasses the cultural elements that help resolve the challenge of selecting between several accounting choices. Societal norms and values influence individuals' conceptions of morality. Individuals' ethical orientation, which is crucial in determining the ethical approach to earnings management, is predominantly influenced by the national culture (Cohen et al., 1993, 1996). Culture delineates the norms and values that are deemed legitimate, acceptable, and desirable (Deephouse et al., 2016; North, 1990). Consequently, people's decisions, understandings, likes, and reasons are shaped by their country's cultures (Chai et al., 2009).

### *2.2 The Relationship Between National Cultural Dimensions and Earnings Management*

Previous research has found empirical evidence supporting a connection between national cultural dimensions (such as collectivism, uncertainty avoidance, masculinity, and power distance) and the practice of earnings management (Bao & Bao, 2004; Callen et al., 2011; Doupnik, 2008; Guan & Pourjalali, 2010; Han et al., 2010; Kanagaretnam et al., 2011; Paredes & Arturo, 2016). Guan & Pourjalali (2010) demonstrated that variations in the selection of discretionary accruals among five Asian nations can be attributed to cultural differences. The study included data from Australia, Japan, Hong Kong, Malaysia, and Singapore. The researchers presented empirical evidence demonstrating that the correlation between earnings management and individualism and uncertainty avoidance exhibited considerable diversity across different characteristics of earnings management. The authors posited that in a culture characterized by high levels of uncertainty, individuals are inclined to mitigate the risk of future performance by employing earnings management techniques that result in decreased income.

Furthermore, cultural dimensions such as individualism, masculinity, and power distance promote a greater degree of autonomy and flexibility for managers in selecting accounting practices. Therefore, these cultural elements have a beneficial influence on earnings management. Furthermore, the author conducted an analysis of enduring societal beliefs regarding earnings manipulation and discovered that variations in these beliefs can account for the fluctuations in activities related to managing wages. The writers contended that the concept of conservatism aligns with enduring social principles over an extended period. Managers who prioritize long-term social values tend to become more cautious and employ strategies to decrease their income through earnings management. Doupnik (2008) examined two aspects of earnings management: earnings smoothing, which involves manipulating accounting accruals, and earnings discretion, which involves manipulating the timing of earnings to reduce their variability.

The study discovered robust and statistically significant correlations between profit management and uncertainty avoidance, as well as individualism. The author conducted a cross-sectional study of thirty-one countries, utilizing Hofstede's cultural dimension measures, individualism, and uncertainty avoidance, as indicators of the national culture of each unique country. The author posited that uncertainty avoidance fosters a tendency among managers to steer clear of fluctuations in wages. This is because individuals in uncertainty avoidance cultures strive to exert control over the unpredictability of future outcomes. Consequently, there is a heightened inclination towards engaging in discretionary earnings management and smoothing. Implementing earnings smoothing strategies provides mutual advantages for all the company's stakeholders. Managers often participate in income-smoothing wages to benefit society as a whole in a society characterized by a low level of individuality. Managers with high power distance have more authority and freedom to choose accounting techniques, leading to increased earnings management. The author stated that there is a correlation between masculinity and monetary success. This correlation motivates managers to participate in earnings management, manipulating earnings to showcase their material achievements.

Han et al. (2010) analyze the cultural value systems and institutional aspects to establish their connections with earnings management. Uncertainty avoidance and individuality were employed as substitutes for the national cultural values system. The researchers discovered that cultural aspects could elucidate managers' decision-making process regarding earnings management. Furthermore, the intensity of investor perception influences the magnitude of these associations.

Kanagaretnam et al. (2011) examined four aspects of national cultures. The study demonstrated that managers in cultures characterized by strong individualism and high masculinity participate in income-increasing earnings management to achieve or surpass the earnings of the previous year. The study utilized data from banks in thirty-nine nations worldwide from 1993 to 2006. The study also found that banks in nations with a high level of individualism, a high level of power distance, and a low level of uncertainty avoidance tend to engage in earnings smoothening. In a culture characterized by high individualism, there is a strong emphasis on personal performance. As a result, corporate managers are driven to meet or exceed the earnings goals. In such a society, managers perceive more significant incentives for undertaking additional risks. Thus, in a society characterized by a strong emphasis on individualism, managers are more likely to participate in earnings management practices aimed at raising income. Based on the research conducted by Gray (1988), the authors of this study believe that in a culture where people vehemently shun uncertainty, they are more likely to choose strict standards and rules that offer little flexibility. In addition, managers in such a society are less likely to engage in tactics that manipulate earnings. Managers have more choices when choosing financial reporting procedures when there is a high power distance. Other members of the firm are less likely to question or dispute their conduct because of the significant power distance. Managers in a society with a considerable power distance are expected to participate in earnings management for opportunistic purposes. Male-dominated cultures emphasize performance-based and assertive behavior (Hofstede & Hofstede, 2001), leading managers in these societies to practice earnings management.

Callen et al. (2011) conducted a study on the cultural dimension of religion and investigated the connection between earnings management and religion and other cultural characteristics. The study findings revealed a lack of correlation between religion and the practice of manipulating financial statements to manipulate earnings. Nevertheless, the study findings indicate that individualism negatively correlates with earnings management, while uncertainty avoidance positively correlates with earnings management.

Zhang et al. (2013) conducted a study on earnings management data and presented compelling empirical proof that supports the correlation between national culture and corporate governance. They demonstrated that there are significant advantages to manipulating financial results to increase revenues in a society that values collective interests over individual interests. In their study, Paredes and Arturo (2016) investigated the impact of various aspects of national culture on natural profit management. The researchers utilized Hofstede's (1980) dataset to assess several aspects of national culture and discovered a favorable correlation between power distance and the practice of natural earnings management. Real earnings management is adversely correlated with individualism, masculinity, and uncertainty avoidance. The empirical research indicates that cultural factors have a substantial impact on the decision-making process of earnings management inside a company. The links between national culture and earnings management vary in direction and strength across different studies. However, it is generally discovered that national culture has a supportive effect on earnings management.

*2.3 The Cultural Effects on The Perceptions and Acceptability of Earnings Management*

Previous studies have investigated how national culture influences individuals' opinions and willingness to accept wage management. The research on the correlation between national culture and earnings management investigates whether there is a connection between national culture and the practice of manipulating financial statements to manipulate reported earnings. The research on the correlation between national culture and earnings management demonstrates the impact of national culture on earnings management. Geiger et al. (2006) investigated the relationship between national cultural variations and the perception of the acceptability of earnings management, as well as its impact on earnings management itself. The study findings revealed significant variations in attitudes and acceptability of wage management based on cultural disparities. The study especially discovered compelling empirical data that demonstrates how power distance and masculine characteristics of culture have a significant impact on individuals' attitudes regarding the acceptability of manipulating accounting earnings. The study results align with previous studies regarding views of earnings manipulation.

Previous research has demonstrated that managers hold varying perspectives regarding the ethical and moral acceptability of earnings management. Earnings management is regarded by certain managers as a matter concerning financial reporting and disclosures. Some individuals view these activities as deceiving users of financial statements, with the intention of benefiting managers at the expense of others. The notion regarding the appropriateness of earnings management is anticipated to impact the extent of earnings management.

Cohen et al. (1993, 1996) posited that national culture plays a role in shaping individuals' ethical orientation, affecting how they perceive and accept earnings management. Hence, empirical data substantiates the claim that national culture influences ethical judgment, and ethical judgment determines the perception of the appropriateness of earnings management.

Lourenço et al. (2018) showed a direct correlation between the perception of corruption and the practice of earnings management in developing nations. The authors established a connection between culture and earnings management through corruption perception. They found that power distance and individualism are positively correlated with corruption perception, based on previous research (Kimbro, 2002). Furthermore, corruption perception is favorably correlated with earnings management.

National cultures exert a substantial influence on the practices of earnings management. Studies suggest that countries with low levels of individualism, high levels of uncertainty avoidance, and long-term orientation are likely to have higher instances of earnings management (Viana Jr et al., 2022; Whelan & Humphries, 2022). Moreover, the correlation between national culture and earnings management differs between developed and emerging economies, with impacts that vary depending on cultural factors such as power distance and masculinity (Mamatzakis et al., 2024). The quality of financial reporting in both Eastern and Western European nations is impacted by cultural aspects, which also influence classification shifting (Reisch, 2021). Although several studies contend that national culture has no direct impact on management's accounting practices, other elements, such as legal requirements and firm-specific qualities, do have an influence (Jarne-Jarne et al., 2022). Moreover, the organizational culture has a significant role in determining the efficacy of corporate governance in overseeing profitability, particularly in relation to factors such as the degree of aversion to uncertainty and the extent of power disparity.

Classification shifting is impacted by the individualistic features of national culture, which also impact innovation, bank leverage, and cultural diversity. Research indicates that nation-states with high levels of individualism also have higher bank leverage (Brewer & Venaik, 2011), and they are also linked to innovation outputs such as creativity and technology (Haq et al., 2018). The individualism-collectivism orientations of different cultures impact management techniques in multinational corporations, making it challenging to transfer domestic practices to overseas companies (Prim et al., 2017).

Furthermore, the psychological constructs at the individual level about the individualism-collectivism aspects may differ throughout countries, exhibiting characteristics such as conformism, ascendancy, and exclusionism (Kulkarni et al., 2010). These results demonstrate the complex interaction between national culture's individualistic aspects and classification changes in different settings. Classification shifting is strongly influenced by the national culture's power distance dimension (Hickey, 2007; Zagladi, 2017). The degree to which people expect and accept hierarchical power distribution is known as power distance (Luptáková et al., 2005). Studies in Banjarmasin, Indonesia, demonstrated a low power distance culture among instructors at private universities, with key indications being the degree of authority exercised by leaders (Sivasubramaniam & Delahunty, 2014). Furthermore, an Australian study discovered that different ethnic groups in the culture had different power-distance values, which affected how they interacted with authoritative figures. Since the power gap influences how people view and react to hierarchical structures, it is essential to comprehend efficient management and communication procedures in businesses. Therefore, in multicultural situations, effective classification shifting requires an understanding of and ability to adjust to power distance differences.

## 3. Data and Methods

*3.1 Research Design*

According to Mc Vay (2006), we calculated a proxy for expected (or average) core earnings for each firm each year using the following model:

$CE_{i,t} = \beta_1 CE_{i,t-1} + \beta_2 ATO_{i,t} + \beta_3 ACCRUALS_{i,t} + \beta_4 ACCRUALS_{i,t-1} + \beta_5 \Delta SALES_{i,t} + \beta_6 NEG\_\Delta SALES_{i,t} + \varepsilon_{i,t}$ ----(1)

The measurements of the variables in the above equation are tabulated below:

| Variables | Definition |
|---|---|
| CE | CE means core earnings calculated as Sales minus Cost of goods sold minus General, selling, and administrative expenses scaled by Sales. |
| ATO | ATO means asset turnover, which is defined as sales divided by average net operating assets. The differences between Operating assets and Operating liabilities calculate the net operating assets.<br>Operating assets = Total asset – Cash and Cash equivalent.<br>Operating liabilities = Total assets – Total debt – Book value of common equity-preferred equity-minority interests. |
| ACCRUALS | ACCUALS mean operating accruals = (Net income before extraordinary item-cash from operation)/Sales |
| ΔSALES | ΔSALES means the percentage change in Sales = $(Sales_t - Sales_{t-1})/Sales_{t-1}$ |
| NEG_ΔSALES | NEG_ΔSALES is a variable showing the percentage change in sales when the change is negative; otherwise, "0". |

Since core earnings are typically persistent, the model incorporates lagged core earnings ($CE_{i,t-1}$). To account for the inverse link between asset turnover and profit margin, asset turnover (ATO) is added. This is crucial, particularly for businesses whose revenue from non-recurring items is high because these businesses are more likely to have adjusted their operational strategy. Lagged accruals ($ACCRUALS_{i,t-1}$) are introduced to capture the information content of last period accruals for current period profits since future performance is correlated with past accruals. By controlling excessive performance resulting from accruals management, current period accruals ($ACCRUALS_t$) enable a robust forecast of abnormal excess of core earnings linked to classification shifting alone. To account for the effect of sales growth on fixed costs—which decrease as sales increase—sales growth ($\Delta SALES_t$) is added. To accommodate varying inclinations for increases and decreases in sales, the model incorporates negative sales (NEG_ΔSALES).

We have calculated the absolute values of the residuals of the regression model 1 as proxies for absolute abnormal core earnings ("Abn_CE"). Then, we have developed the following regression model to calculate the effects of different dimensions of national culture (Individualism, power distance, uncertainty avoidance, and masculinity):

$Abn\_CE = \beta_0 + \beta_2 IDV + \beta_3 PDI + \beta_4 UAI + \beta_5 MAS + \beta_{6-8} CONTROLS + Year\_FE + Indusry\_FE + \mu$ ----(2)

Where "*IDV*," "PDI," "UAI," and *"MAS"* are the indices of four Hofstede cultural dimensions: individualism, power distance, uncertainty avoidance, and masculinity, respectively. The model includes firm size *("SIZE"),* return on assets *("ROA"),* and Leverage *("LEVERAGE")* as control variables. *Year_FE* and *Indusry_FE* are year-fixed effects and industry-fixed effects, respectively. *μ* represents the error terms.

*3.2 Sample*

Study data was collected from the Compustat global database and the Hofstede database. The study sample comprises 77,737 firm-year observations from 2018 to 2022 in 35 countries. Sample firms in a country are assumed to be representative of the country's culture, language, and geography. Initial data was cleaned for missing values, and the final sample consisted of 77,737 firm-year observations. Table 1 and Table 2 represent descriptive statistics of the variables.

Table 1. Descriptive Statistics of variables

| Variables | Obs | Mean | Std. Dev. | Min | Max | p1 | p99 | Skew. | Kurt. |
|---|---|---|---|---|---|---|---|---|---|
| Abn_CE | 7773 | 5.714 | 152.442 | 0.000 | 18809.094 | 0.001 | 37.825 | 63.960 | 5409.329 |
| PDI | 7773 | 60.403 | 18.644 | 11.000 | 104.000 | 22.000 | 104.00 | 0.129 | 2.945 |
| IDV | 7773 | 46.210 | 22.715 | 12.000 | 90.000 | 14.000 | 90.000 | 0.455 | 2.228 |
| MAS | 7773 | 57.548 | 22.828 | 5.000 | 95.000 | 5.000 | 95.000 | 0.038 | 2.979 |
| UAI | 7773 | 59.649 | 25.683 | 8.000 | 112.000 | 8.000 | 94.000 | 0.043 | 1.623 |
| SIZE | 7773 | 8.186 | 3.294 | 1.567 | 15.994 | 1.567 | 15.994 | 0.181 | 2.335 |
| ROA | 7773 | -0.003 | 0.154 | -0.850 | 0.266 | -0.850 | 0.266 | -2.982 | 14.93 |
| LEVERAGE | 7773 | 0.454 | 0.220 | -0.004 | 1.000 | 0.027 | 0.940 | 0.088 | 2.300 |

Note: *Abn_CE* = Abnormal core earnings, *PDI* = Power distance index, *IND* = Individualism/collectivism index, *MAS* = Masculinity index, *UAI* = Uncertainty avoidance index, *SIZE* = size of the business measured by the natural logarithm of total assets, *ROA* = Return on assets, *LEVERAGE* = Leverage ratio.

Table 2. Variable means by country

|    | Countries      | Abs_CE | PDI | IDV | MAS | UAI | SIZE   | ROA    | LEVERAGE |
|----|----------------|--------|-----|-----|-----|-----|--------|--------|----------|
| 1  | Argentina      | 0.126  | 49  | 46  | 56  | 86  | 9.603  | 0.022  | 0.548    |
| 2  | Australia      | 49.960 | 38  | 90  | 61  | 51  | 4.300  | -0.160 | 0.371    |
| 3  | Austria        | 0.096  | 11  | 55  | 79  | 70  | 6.811  | 0.023  | 0.590    |
| 4  | Belgium        | 6.843  | 65  | 75  | 54  | 94  | 6.463  | 0.007  | 0.537    |
| 5  | Brazil         | 0.859  | 69  | 38  | 49  | 76  | 8.290  | 0.037  | 0.576    |
| 6  | Chile          | 5.334  | 63  | 23  | 28  | 86  | 10.767 | 0.030  | 0.509    |
| 7  | Colombia       | 0.412  | 67  | 13  | 64  | 80  | 13.803 | 0.030  | 0.469    |
| 8  | Denmark        | 17.135 | 18  | 74  | 16  | 23  | 6.670  | -0.049 | 0.481    |
| 9  | Finland        | 0.438  | 33  | 63  | 26  | 59  | 5.354  | 0.019  | 0.528    |
| 10 | France         | 5.153  | 68  | 71  | 43  | 86  | 5.960  | -0.015 | 0.581    |
| 11 | Germany        | 0.644  | 35  | 67  | 66  | 65  | 5.928  | 0.003  | 0.554    |
| 12 | Great Britain  | 2.752  | 35  | 89  | 66  | 35  | 5.119  | -0.031 | 0.471    |
| 13 | Greece         | 0.163  | 60  | 35  | 57  | 112 | 4.583  | 0.020  | 0.583    |
| 14 | Hong Kong      | 2.146  | 68  | 25  | 57  | 29  | 7.038  | -0.032 | 0.430    |
| 15 | India          | 2.674  | 77  | 48  | 56  | 40  | 7.833  | 0.036  | 0.441    |
| 16 | Indonesia      | 0.414  | 78  | 14  | 46  | 48  | 12.793 | 0.023  | 0.451    |
| 17 | Ireland        | 1.403  | 28  | 70  | 68  | 35  | 6.668  | -0.012 | 0.467    |
| 18 | Italy          | 0.617  | 50  | 76  | 70  | 75  | 5.537  | 0.016  | 0.596    |
| 19 | Japan          | 0.328  | 54  | 46  | 95  | 92  | 10.658 | 0.028  | 0.459    |
| 20 | Korea South    | 0.701  | 60  | 18  | 39  | 85  | 12.587 | -0.004 | 0.431    |
| 21 | Malaysia       | 0.241  | 104 | 26  | 50  | 36  | 5.926  | 0.008  | 0.367    |
| 22 | Mexico         | 0.112  | 81  | 30  | 69  | 82  | 10.332 | 0.036  | 0.543    |
| 23 | Netherlands    | 8.968  | 38  | 80  | 14  | 53  | 6.889  | -0.002 | 0.565    |
| 24 | New Zealand    | 0.460  | 22  | 79  | 58  | 49  | 5.537  | -0.024 | 0.467    |
| 25 | Norway         | 26.550 | 31  | 69  | 8   | 50  | 7.218  | -0.041 | 0.514    |
| 26 | Peru           | 0.460  | 64  | 16  | 42  | 87  | 6.977  | 0.040  | 0.485    |
| 27 | Philippines    | 0.528  | 94  | 32  | 64  | 44  | 9.542  | 0.023  | 0.439    |
| 28 | Portugal       | 0.113  | 63  | 27  | 31  | 104 | 6.374  | 0.025  | 0.630    |
| 29 | Singapore      | 0.679  | 74  | 20  | 48  | 8   | 5.193  | -0.014 | 0.439    |
| 30 | Spain          | 0.153  | 57  | 51  | 42  | 86  | 6.293  | 0.025  | 0.607    |
| 31 | Sweden         | 27.138 | 31  | 71  | 5   | 29  | 5.911  | -0.136 | 0.435    |
| 32 | Switzerland    | 18.979 | 34  | 68  | 70  | 58  | 6.866  | 0.026  | 0.496    |
| 33 | Thailand       | 2.090  | 64  | 20  | 34  | 64  | 8.244  | 0.032  | 0.406    |
| 34 | Turkey         | 0.386  | 66  | 37  | 45  | 85  | 6.869  | 0.076  | 0.513    |
| 35 | Venezuela      | 15.749 | 81  | 12  | 73  | 76  | 10.758 | 0.078  | 0.399    |

Figure 1. Hofstede cultural dimensions across countries

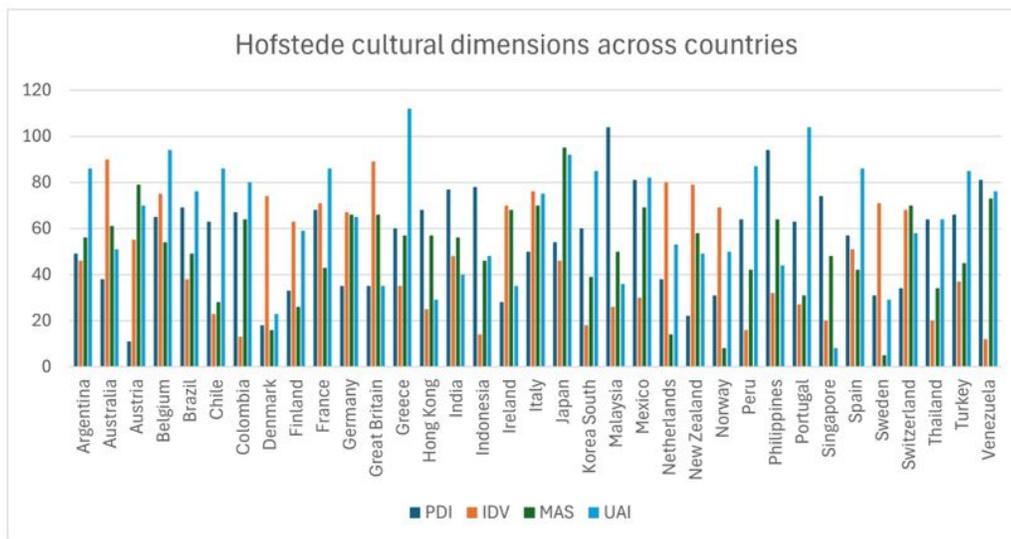

The correlation among variables shows that the individualism dimension of national culture positively correlates with the Abnormal core earnings. However, Power distance, masculinity, and uncertainty avoidance are negatively correlated with Abnormal core earnings. Firm size, return on assets, and Leverage are negatively correlated with the Abnormal core earnings.

Table 3. Correlation among variables

| Variables | (1) | (2) | (3) | (4) | (5) | (6) | (7) | (8) |
|---|---|---|---|---|---|---|---|---|
| (1) Abn CE | 1.000 | | | | | | | |
| (2) PDI | -0.037*** | 1.000 | | | | | | |
| (3) IDV | 0.048*** | -0.647*** | 1.000 | | | | | |
| (4) MAS | -0.019*** | -0.024*** | 0.106*** | 1.000 | | | | |
| (5) UAI | -0.018*** | -0.190*** | -0.125*** | 0.433*** | 1.000 | | | |
| (6) SIZE | -0.047*** | 0.123*** | -0.478*** | 0.186*** | 0.472*** | 1.000 | | |
| (7) ROA | -0.090*** | 0.178*** | -0.178*** | 0.118*** | 0.118*** | 0.298*** | 1.000 | |
| (8) LEVERAGE | -0.040*** | -0.069*** | 0.055*** | 0.013*** | 0.092*** | 0.119*** | -0.050*** | 1.000 |

Note: *** $p < 0.01$, ** $p < 0.05$, * $p < 0.10$

## 4. Data Analysis and Results

Table 4 represents the data analysis results of the OLS regression for abnormal core earnings. The independent variables are the four Hofstede cultural dimensions: Power distance (PDI), Individualism (IDV), Masculinity (MAS), and uncertainty avoidance. In addition, firm size, return on assets (ROA), and Leverage are used as control variables.

Table 4. Regression analysis dependent variable: Abnormal core earnings ("Abn_CE")

| Variables | Abn_CE |
|---|---|
| PDI | -0.0333 |
|  | (0.0443) |
| IDV | 0.161*** |
|  | (0.0405) |
| MAS | -0.104*** |
|  | (0.0286) |
| UAI | 0.0515* |
|  | (0.0283) |
| SIZE | 0.0346 |
|  | (0.233) |
| ROA | -77.39*** |
|  | (3.809) |
| LEVERAGE | -25.24*** |
|  | (2.616) |
| Constant | 14.15*** |
|  | (5.229) |
| Observations | 77,737 |
| R-squared | 0.018 |
| Year FE & Industry FE | Yes |

Note: Standard errors in parentheses;

*** $p<0.01$, ** $p<0.05$, * $p<0.10$

The relationship between power distance and abnormal core earnings is non-significant, which implies that we cannot reject the null hypothesis that power distance does not influence abnormal core earnings. The relationship between individualism and abnormal core earnings is positive and statistically significant at a 1% significance level. It implies that the higher the value of the individualism index, the higher the abnormal core earnings. Countries characterized by high individualism encourage business managers to engage in core earnings management. In individualistic culture, individuals are motivated by personal interests. The relationship between masculinity and abnormal earnings is negative and statistically significant at a 1% significance level. It implies that masculinity reduces abnormal core earnings in an organization. The result is consistent with the view that masculinity is correlated with risk taking behavior. Managers in highly masculine countries need more incentives to engage in core earnings management. The relationship between uncertainty avoidance and abnormal core earnings is positive and statistically significant at a 10% significance level. This implies that the higher the uncertainty avoidance value, the higher the core earnings management. Managers in countries with high uncertainly avoidance tend to avoid fluctuations in core earnings and engage in core earnings management.

The study analysis controlled the time-variant yearly effect by incorporating year-fixed effects in the model. It also controlled for time-invariant industry effects on abnormal core earnings.

## 5. Practical Implications

Gaining insight into the impact of cultural factors on profit management can help multinational organizations develop governance practices that are attuned to the cultural contexts of their different operations. Auditors can customize their methods and level of doubt based on the cultural tendency to manipulate earnings, enhancing the efficiency and precision of financial audits in various cultural contexts. Regulators should contemplate incorporating cultural assessments into their regulatory frameworks, which could provide culturally adapted rules that effectively eliminate unethical financial reporting activities.

## 6. Suggestions for Further Research

The results of this study offer a detailed comprehension of how Hofstede's cultural aspects impact aberrant core earnings management in different national cultures. Nevertheless, they also create opportunities for additional investigation and study in global business and accounting. Here are several recommendations for future research.

First, although Hofstede's dimensions provide valuable insights into the impact of culture on profit management, a more thorough analysis can be achieved by integrating additional cultural frameworks such as the GLOBE research, Trompenaars' Dimensions, or Schwartz's Value Orientations. Examining and contrasting outcomes across diverse cultural frameworks might deepen our comprehension of how culture influences accounting methodologies. Second, this study employed a cross-sectional methodology, which allows for a momentary assessment based on preexisting data. Subsequent studies could utilize longitudinal approaches to monitor the evolution of revenue manipulation strategies over time within the same cultural environments. Examining the relationship between shifts in cultural factors and changes in accounting behaviors might enhance our understanding of any potential correlation across time. Third, further refinement of the research could be achieved by incorporating additional control factors such as market conditions, corporate governance frameworks, or conformity with international financial reporting requirements. These factors may also impact the correlation between cultural characteristics and earnings management.

In addition to the quantitative study, qualitative research methodologies such as case studies, interviews, or ethnographic studies could offer more profound insights into how culture impacts managerial decision-making in earnings management. Various industries may respond in distinct ways to cultural influences. Conducting sector-specific studies can determine whether industries are more susceptible to cultural influences on profit management than others. A more detailed understanding of cultural impacts can be obtained by conducting regional or country-specific studies. This research could assist in identifying distinct cultural groups where the impact on earnings management is more noticeable.

## 7. Conclusion

Extant literature examined multiple dimensions of the relationship between national culture and earnings management and provided strong empirical support to favor the relationship. Variability in earnings management is associated with variability in national culture. In addition to the direct impact of national culture on earnings management, cultural variability is responsible for developing variable perceptions about the acceptability of earnings management. Extant literature also shows the effect of national culture on interaction. The relationships of earnings management with corruption and the adoption of accounting standards vary depending on the cultural variability of the countries studied. Prior research is needed to provide more information to argue that earnings management is a product of culture; instead, it may be claimed with empirical support that earnings management is strongly associated with culture.

Overall, this study theoretically enhances our comprehension of the significant impact that deeply rooted cultural characteristics have on business ethics and practices, particularly in relation to the manipulation of financial earnings. The research highlights the importance of considering cultural elements when developing a global company strategy, audit processes, and regulatory laws. As firms increasingly grow internationally, it is essential to include cultural understanding into the corporate and regulatory framework to maintain ethical standards and accountability on a global scale.

## References

Ball, R., & Brown, P. (2013). An empirical evaluation of accounting income numbers. In *Financial Accounting and Equity Markets* (pp. 27–46). Routledge.

Bao, B.-H., & Bao, D.-H. (2004). Income smoothing, earnings quality and firm valuation. *Journal of Business Finance & Accounting*, *31*(9-10), 1525–1557.

Beaver, W. H., Correia, M., & McNichols, M. F. (2012). Do differences in financial reporting attributes impair the predictive ability of financial ratios for bankruptcy? *Review of Accounting Studies*, *17*, 969–1010.

Brewer, P., & Venaik, S. (2011). *Individualism–collectivism in Hofstede and GLOBE*. Springer.

Callen, J. L., Morel, M., & Richardson, G. (2011). *Do culture and religion mitigate earnings management? Evidence from a cross-country analysis*. https://doi.org/10.1057/JDG.2010.31

Chai, S.-K., Liu, M., & Kim, M.-S. (2009). Cultural comparisons of beliefs and values: Applying the grid-group approach to the World Values Survey. *Beliefs and Values*, *1*(2), 193–208.

Cohen, J. R., Pant, L. W., & Sharp, D. J. (1993). Culture-based ethical conflicts confronting multinational accounting firms. *Accounting Horizons*, *7*(3), 1.

Cohen, J. R., Pant, L. W., & Sharp, D. J. (1996). A methodological note on cross-cultural accounting ethics research. *The International Journal of Accounting*, *31*(1), 55–66.

Deephouse, D. L., Newburry, W., & Soleimani, A. (2016). The effects of institutional development and national culture on cross-national differences in corporate reputation. *Journal of World Business*, *51*(3), 463–473.

Doupnik, T. (2008). *Influence of Culture on Earnings Management: A Note*. https://doi.org/10.1111/j.1467-6281.2008.00265.x

Francis, J., LaFond, R., Olsson, P. M., & Schipper, K. (2004). Costs of equity and earnings attributes. *The Accounting Review*, *79*(4), 967–1010.

Geiger, M. A., O'Connell, B. T., Clikeman, P. M., Ochoa, E., Witkowski, K., & Basioudis, I. (2006). Perceptions of earnings management: The effects of national culture. *Advances in International Accounting*, *19*, 175–199.

Gray, S. J. (1988). Towards a theory of cultural influence on the development of accounting systems internationally. *Abacus*, *24*(1), 1–15.

Guan, L., & Pourjalali, H. (2010). Effect of cultural environmental and accounting regulation on earnings management: A multiple year-country analysis. *Asia-Pacific Journal of Accounting & Economics*, *17*(2), 99–127.

Han, S., Kang, T., Salter, S. B., & Yoo, Y. (2010). *A cross-country study on the effects of national culture on earnings management*. https://doi.org/10.1057/JIBS.2008.78

Haq, M., Hu, D., Faff, R., & Pathan, S. (2018). New evidence on national culture and bank capital structure. *Pacific-Basin Finance Journal*, *50*, 41–64.

Healy, P. M., & Wahlen, J. M. (1999). A review of the earnings management literature and its implications for standard setting. *Accounting Horizons*, *13*(4), 365–383.

Hickey, S. (2007). *Power Distance in Cross Cultural Workplaces.* Dublin, National College of Ireland.

Hofstede, G. H., & Hofstede, G. (2001). *Culture's consequences: Comparing values, behaviors, institutions and organizations across nations*. sage.

Jarne-Jarne, J. I., Callao-Gastón, S., Marco-Fondevila, M., & Llena-Macarulla, F. (2022). The Impact of Organizational Culture on the Effectiveness of Corporate Governance to Control Earnings Management. *Journal of Risk and Financial Management*, *15*(9), 379.

Kanagaretnam, K., Lim, C., & Lobo, G. J. (2011). *Effects of national culture on earnings quality of banks*. https://doi.org/10.2139/ssrn.1734014

Kimbro, M. B. (2002). A cross-country empirical investigation of corruption and its relationship to economic, cultural, and monitoring institutions: An examination of the role of accounting and financial statements quality. *Journal of Accounting, Auditing & Finance*, *17*(4), 325–350.

Kothari, S. P. (2001). Capital markets research in accounting. *Journal of Accounting and Economics*, *31*(1–3), 105–231.

Kulkarni, S. P., Hudson, T., Ramamoorthy, N., Marchev, A., Georgieva-Kondakova, P., & Gorskov, V. (2010). Dimensions of individualism-collectivism. *Verslo Ir Teisės Aktualijos*, *93*.

Leuz, C., Nanda, D., & Wysocki, P. D. (2003). Earnings management and investor protection: An international comparison. *Journal of Financial Economics*, *69*(3), 505–527.

Lopez, T. J., & Rees, L. (2002). The effect of beating and missing analysts' forecasts on the information content of unexpected earnings. *Journal of Accounting, Auditing & Finance*, *17*(2), 155–184.

Lourenço, I. C., Rathke, A., Santana, V., & Branco, M. C. (2018). Corruption and earnings management in developed and emerging countries. *Corporate Governance: The International Journal of Business in Society*, *18*(1), 35–51.

Luptáková, S., Vargic, B., & Kincel, I. (2005). National culture dimension of power distance in the Baltic States. *Journal of Business Economics and Management*, *6*(2), 61–69.

Mamatzakis, E. C., Neri, L., & Russo, A. (2024). Do cultural differences affect the quality of financial reporting in the EU? An analysis of Western EU countries vis a vis Eastern EU countries. *Journal of Accounting & Organizational Change*, *20*(2), 248–275.

North, D. C. (1990). *Institutions, institutional change and economic performance*.

Paredes, P., & Arturo, Á. (2016). *The association of real earnings management with: Enterprise resource planning systems, audit effort, and future financial performance*. https://doi.org/10.25148/ETD.FIDC000701

Prim, A. L., FILHO, L. S., Zamur, G. A. C., & Di Serio, L. C. (2017). The relationship between national culture dimensions and degree of innovation. *International Journal of Innovation Management*, *21*(01), 1730001.

Reisch, L. (2021). Does national culture influence management's accounting behaviour and strategy?–an empirical analysis of European IFRS adopters. *Cross Cultural & Strategic Management*, *28*(1), 129–157.

Sivasubramaniam, D., & Delahunty, J. (2014). Cultural variation in Australia: Ethnicity, host community residence, and power-distance values. *Cross-Cultural Communication*, *10*(4), 136–144.

Viana Jr, D. B. C., Lourenço, I. M. E. C., Ohlson, M., & Augusto SF de Lima, G. (2022). National culture and earnings management in developed and emerging countries. *Journal of Accounting in Emerging Economies*, *12*(1), 150–186.

Whelan, C., & Humphries, S. A. (2022). Examining the Relationship Between National Culture and Earnings Management. *Journal of Applied Business & Economics*, *24*(6).

Zagladi, A. N. (2017). Power distance as a national culture observed in organizational scope. *Journal of Management and Marketing Review (JMMR) Vol*, *2*(3).

Zhang, X., Liang, X., & Sun, H. (2013). Individualism–collectivism, private benefits of control, and earnings management: A cross-culture comparison. *Journal of Business Ethics*, *114*, 655–664.

Zucker, L. G. (1983). Organizations as institutions. *Research in the Sociology of Organizations*, *2*(1), 1–47.